\documentclass[superscriptaddress,12pt]{revtex4}
\usepackage{amsfonts}
\usepackage{graphicx}
\usepackage{marvosym}

\begin{document}
\title{Generation of genuine $\chi$-type four-particle entangled state of  superconducting artificial atoms with broken symmetry {\footnote{Chun-Ling Leng $\cdot$ Xin Ji ({\large \Letter})  $\cdot$ Shou Zhang\\
 Department of Physics, College of Science, Yanbian University,
\\ Yanji, Jilin 133002, People's Republic of China \\ Qi Guo $\cdot$ Condensed Matter Science and Technology Institute, Department of Physics, Harbin Institute of Technology,
Harbin, Heilongjiang 150001, China
\\
 e-mail: jixin@ybu.edu.cn}}}
\author{\textbf{Chun-Ling Leng $\cdot$ Qi Guo $\cdot$  Xin Ji $\cdot$ Shou Zhang} }

\begin{abstract}
\noindent \textbf{{Abstract}}:  We propose a scheme for generating
a genuine $\chi$-type four-particle entangled state of
superconducting artificial atoms with broken symmetry by using
one-dimensional transmission line resonator as a data bus. The
$\Delta$-type three-level artificial atom we use in the scheme is
different from natural atom and has cyclic transitions. After
suitable interaction time and simple operations, the desired
entangled state can be obtained. Since artificial atomic excited
states and photonic states are adiabatically eliminated, our
scheme is robust against the spontaneous emissions of artificial
atoms and the decays of transmission line resonator.
\end{abstract}

\maketitle
\section*{1. Introduction}

Quantum entanglement, a remarkable and attractive feature of
quantum mechanics, plays a significant role not only in testing
quantum nonlocality, but also in processing a variety of quantum
information tasks \cite{AKE,CHBS,KMH,SBZGCG,GV}. Therefore,
preparation of various quantum entangled states has been being an
important subject in quantum information science since a few
decades ago \cite{SB,GCGY,PBSO,DLEK,JCH,XBZ,DBH,XQS,HYYL,XYCPY}.
Multi-particle entangled states, such as GHZ state \cite{GHZ1989},
W state \cite{W2000}, cluster state \cite{BHJ2001}, etc, are the
fundamental resource of quantum information processing. In the
research of faithful teleportation of an arbitrary two-qubit state
with multipartite entanglement, Yeo and Chua \cite{YY2006}
introduced a genuine four-qubit entangled state
$|\chi^{00}\rangle_{3214}$, namely $\chi$-type entangled state
\begin{eqnarray}\label{01}
|\chi^{00}\rangle_{3214}=\frac{1}{2\sqrt{2}}(|0000\rangle-|0011\rangle-|0101\rangle+|0110\rangle
+|1001\rangle+|1010\rangle+|1100\rangle+|1111\rangle)_{3214},
\end{eqnarray}
which can't be transformed into other types of multipartite
entangled states by stochastic local operations and classical
communication. The state $|\chi^{00}\rangle_{3214}$ has many
interesting entanglement properties. It has been shown that a new
Bell inequality is optimally violated by
$|\chi^{00}\rangle_{3214}$ but not by other types of entangled
states \cite{WCF2007}. Another important property is that it has
the maximum entanglement between qubits $(3,2)$ and $(1,4)$, and
between $(3,1)$ and $(2,4)$. Furthermore, $\chi$-type entangled
state also has many applications in quantum dense coding
\cite{YY2006} or in quantum secure direct communication
\cite{SL2008}. Thus, recent researches have focused on looking for
some good methods and systems for generating $\chi$-type entangled
state. For example, Yuuki \emph{et al}. \cite{YT2005} and Wang
\emph{et al}. \cite{WHF2009} separately proposed a scheme for
generating $\chi$-type four-photon entangled state by using liner
optics elements and photon detectors. Wang \emph{et al}.
\cite{WXW2008} and Shi \emph{et al}. \cite{SYL2012} respectively
introduced an approach to realize $|\chi^{00}\rangle_{3214}$-like
state and $|\chi^{00}\rangle_{3214}$ state in ion trap system.
Zhang \emph{et al}. \cite{ZYJ2009} proposed a scheme to generate
$|\chi^{00}\rangle_{3214}$ state of four atomic qubits via two
Bell states in cavity quantum electrodynamics system (QED).
Afterwards, Guo \cite{G2012} \emph{et al}. proposed a scheme to
generate $\chi$-type entangled state in
 circuit QED system.

There are many physical carriers of quantum information, such as
photons, atoms and ions, and so on. Among those physical carriers,
 superconducting artificial atoms (AAs) attract lots of attentions because
they have potentially excellent scalability due to
well-established microfabrication techniques. In recent years, a
new system called circuit QED has been put forward, in which
superconducting AAs are fabricated inside transmission line
resonator (TLR). Circuit QED system is analogous to cavity QED, in
which superconducting AAs play the role of atoms in cavity QED
system. This architecture is considered as a promising candidates
for implementing quantum information processing
\cite{YM2001,YJQ2011,LD2010} because it strongly inhibits
spontaneous emission \cite{HAA2008}, allows high-fidelity quantum
nondemolition measurements of multiple qubit states \cite{GJ2006},
and can couple qubits separated by centimeter-scale distances
\cite{MJ2007,DCL2009}.

In this paper, we propose a new scheme to create genuine
$\chi$-type entangled state of superconducting AAs. It is
generally known that artificial solid-state atoms with long-lived
internal states are suitable for the storage of quantum
information. In addition, our scheme can obtain both the
controllable and selective interqubit coupling, meanwhile the
operations are simple, which is helpful to the scalable quantum
information processing. We encode the quantum information on the
ground states, which can greatly reduce the effect of artificial
atomic spontaneous emission on the results, and we can generate
the $\chi$-type four-particle entangled state with high fidelity.
This paper is arranged as follows. Firstly, the fundamental model
and the Hamiltonian of the system are introduced. Then, we show
how to prepare the $\chi$-type four-particle entangled state of
superconducting AAs with broken symmetry in one-dimensional TLR.
Finally, we analyze the performance and the experimental
feasibility of the scheme, and give a conclusion.

\section*{2. The fundamental model and Hamiltonian}

\begin{figure}[htp]
\begin{center}
  \includegraphics[width=10cm]{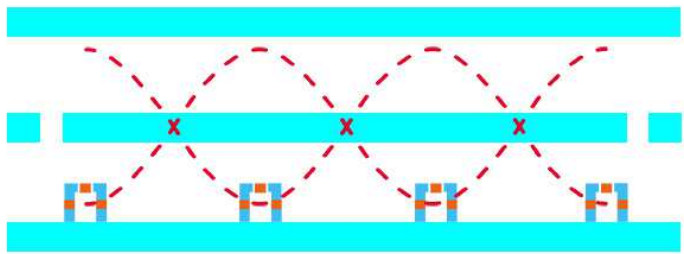}\\
  \caption{The schematic
setup for generating genuine $\chi$-type four-particle entangled
state. Four $\Delta$-type three-level artificial atoms are
fabricated inside a one-dimensional transmission line resonator.
}\label{fig1}
\end{center}
\end{figure}

\begin{figure}[htp]
\begin{center}
  \includegraphics[width=7cm]{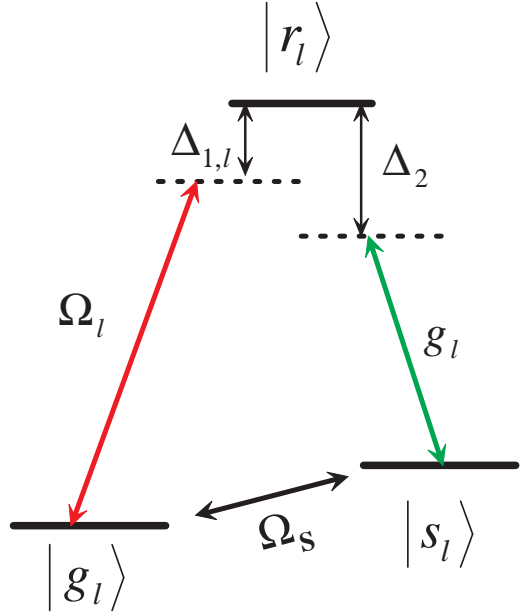}\\
  \caption{The level
configuration of $\Delta$-type three-level artificial
atom.}\label{fig2}
\end{center}
\end{figure}

The schematic setup for generating genuine $\chi$-type entangled
state of AAs is shown in Fig.~\ref{fig1}.
 There are four three-level AAs
fabricated inside a one-dimensional TLR. The level configuration
of the AA interacting with TLR is shown in Fig.~\ref{fig2}. The
states $|g_{l}\rangle$ and $|s_{l}\rangle$ are two ground levels
and $|r_{l}\rangle$ is an excited level of the AAs. This kind of
``artificial atoms" can be made of superconducting three-junction
flux qubit circuits \cite{LYX2005}. For natural three-level atoms,
because of the optical selection rule, the optical transition
between the lowest two levels is forbidden. In contrast to the
natural atoms, the potential energies for superconducting flux
qubit circuits can be artificially controlled. When the reduced
bias magnetic flux $f=\Phi/\Phi_{0}=1/2$, where $\Phi$ is the
static magnetic flux applied to the loop and $\Phi_{0}$ is the
magnetic-flux quantum, artificial atom owns the parity symmetries,
and the transitions of lowest three levels behave like a
$\Lambda$-type or ladder-type natural atom. In this case, the
dipole transition $|g_{l}\rangle\leftrightarrow|s_{l}\rangle$ is
forbidden, while the other two transitions
$|g_{l}\rangle\leftrightarrow|r_{l}\rangle$ and
$|s_{l}\rangle\leftrightarrow|r_{l}\rangle$ are allowed. However,
when $f\neq1/2$, the parity symmetry is broken for the interaction
Hamiltonian. Therefore, all three dipole transitions among
$|g_{l}\rangle\leftrightarrow|s_{l}\rangle$,
$|s_{l}\rangle\leftrightarrow|r_{l}\rangle$ and
$|r_{l}\rangle\leftrightarrow|g_{l}\rangle$ are possible, allowing
the artificial atom to be $\Delta$-type\cite{YJQ2011}.

The transition $|g_{l}\rangle\leftrightarrow|r_{l}\rangle$ of
qubit $l$ is driven by the classical field with Rabi frequency
$\Omega_{l}$. The transition
$|s_{l}\rangle\leftrightarrow|r_{l}\rangle$ of qubit $l$ is
coupled to TLR with the coupling constant $g_{l}$. The frequency
detunings between the artificial atomic transitions
$|g_{l}\rangle\leftrightarrow|r_{l}\rangle,
|s_{l}\rangle\leftrightarrow|r_{l}\rangle$ and the relevant
classical field and TLR are denoted as $\Delta_{1,l}$ and
$\Delta_{2}$, respectively. In the interaction picture, the
Hamiltonian describing the AAs-field interaction\cite{SCP2005} is
\begin{eqnarray}\label{03}
H_{\mathrm{int}}&=&\sum_{l=1}^{4}(e^{i\Delta_{1,l}t}\Omega_{l}|r_{l}\rangle\langle
g_{l}|+g_{l}ae^{i\Delta_{2}t}|r_{l}\rangle\langle
s_{l}|\cr&&+\Omega_{S}|s_{l}\rangle\langle g_{l}|+\mathrm {H.c.}).
\end{eqnarray}
For the sake of simplicity, we assume $g_{1} = g_{2} = g_{3} =
g_{4} = g$ in this paper.

Under the conditions $\Delta_{1,l}\gg\Omega_{l}, \Delta_{2}\gg g$,
the upper level $|r_{l}\rangle$ in the Hamiltonian
$H_{\mathrm{int}}$ can be adiabatically eliminated, leading to the
couplings between the two ground states
\begin{eqnarray}\label{09}
H_{\mathrm{int}}^{'}&=&-\sum_{l=1}^{4}[\eta_{l}|g_{l}\rangle\langle
g_{l}| +\xi a^{\dagger}a|s_{l}\rangle\langle s_{l}|\cr
&&+\lambda_{l}(a^{\dagger}S^{+}_{l}e^{-i\delta_{l}t}+\mathrm
{H.c.})-(\Omega_{S}S^{+}_{l}+\mathrm {H.c.})],
\end{eqnarray}
where

\begin{eqnarray}
\eta_{l}&=&\frac{\Omega_{l}^{2}}{\Delta_{1,l}},\nonumber\\
\xi&=&\frac{g^{2}}{\Delta_{2}},\nonumber\\
\lambda_{l}&=&\frac{\Omega_{l}
g}{2}(\Delta_{1,l}^{-1}+\Delta_{2}^{-1}),\nonumber\\
\delta_{l}&=&\Delta_{2}-\Delta_{1,l},\nonumber\\
S^{+}_{l}&=&|s_{l}\rangle\langle g_{l}|,~
S^{-}_{l}=|g_{l}\rangle\langle s_{l}|.
\end{eqnarray}

Next, we define the new basis
\begin{eqnarray}\label{10}
|+_{l}\rangle=\frac{1}{\sqrt{2}}(|g_{l}\rangle+|s_{l}\rangle),~
|-_{l}\rangle=\frac{1}{\sqrt{2}}(|g_{l}\rangle-|s_{l}\rangle),
\end{eqnarray}

The Hamiltonian rewritten in new basis is
\begin{eqnarray}\label{11}
H^{\mathrm{new}}&=&-\sum_{l=1}^{4}\{\frac{\eta_{l}}{2}(\sigma^{+}_{l}\sigma^{-}_{l}+\sigma^{-}_{l}\sigma^{+}_{l}
+\sigma^{+}_{l}+\sigma^{-}_{l})\cr
&&+\frac{\xi}{2}a^{\dagger}a(\sigma^{+}_{l}\sigma^{-}_{l}+\sigma^{-}_{l}\sigma^{+}_{l}
-\sigma^{+}_{l}-\sigma^{-}_{l})\cr
&&+\lambda_{l}e^{i\delta_{l}t}a(\sigma_{z,l}+\frac{1}{2}\sigma^{-}_{l}-\frac{1}{2}\sigma^{
+}_{l})\cr
&&+\lambda_{l}e^{-i\delta_{l}t}a^{\dagger}(\sigma_{z,l}+\frac{1}{2}\sigma^{+}_{l}-\frac{1}{2}\sigma^{
-}_{l})\cr&&-2\Omega_{S}\sigma_{z,l}\},
\end{eqnarray}
where
$\sigma_{z,l}=\frac{1}{2}(|+_{l}\rangle\langle+_{l}|-|-_{l}\rangle\langle-_{l}|)$,
$\sigma^{+}_{l}=|+_{l}\rangle\langle-_{l}|$ and
$\sigma^{-}_{l}=|-_{l}\rangle\langle+_{l}|$.

 In the interaction picture with reference to
$H^{\mathrm{new}}_{\mathrm{0}}=\sum_{l=1}^{4}2\Omega_{S}\sigma_{z,l}$,
the Hamiltonian is reduced to
\begin{eqnarray}\label{12}
&H^{\mathrm{new}}_{\mathrm{int}}&=-\sum_{l=1}^{4}\{\frac{\eta_{l}}{2}(e^{i\Omega_{S}t}\sigma^{+}_{l}\sigma^{-}_{l}+e^{-i\Omega_{S}t}\sigma^{-}_{l}\sigma^{+}_{l}\cr&&
+e^{i2\Omega_{S}t}\sigma^{+}_{l}+e^{-i2\Omega_{S}t}\sigma^{-}_{l})\cr
&&+\frac{\xi}{2}a^{\dagger}a(e^{i\Omega_{S}t}\sigma^{+}_{l}\sigma^{-}_{l}+e^{-i\Omega_{S}t}\sigma^{-}_{l}\sigma^{+}_{l}\cr&&
-e^{i2\Omega_{S}t}\sigma^{+}_{l}-e^{-i2\Omega_{S}t}\sigma^{-}_{l})\cr
&&+\lambda_{l}e^{-i\delta_{l}t}a^{\dagger}(\sigma_{z,l}+\frac{1}{2}e^{-i2\Omega_{S}t}\sigma^{-}_{l}-\frac{1}{2}e^{i2\Omega_{S}t}\sigma^{
+}_{l})\cr
&&+\lambda_{l}e^{i\delta_{l}t}a(\sigma_{z,l}+\frac{1}{2}e^{i2\Omega_{S}t}\sigma^{+}_{l}-\frac{1}{2}e^{-i2\Omega_{S}t}\sigma^{
-}_{l})\}.
\end{eqnarray}

Assume that 2$\Omega_{S}\gg\delta_{l}, \lambda_{l},\eta_{l},\xi$,
we can neglect the terms oscillating fast. Then
$H^{\mathrm{new}}_{\mathrm{int}}$ reduces to
\begin{eqnarray}\label{13}
H^{\mathrm{new}}_{\mathrm{int}}=-\sum_{l=1}^{4}\frac{\lambda_{l}}{2}(e^{i\delta_{l}t}a+e^{-i\delta_{l}t}a^{\dagger})(S^{+}_{l}+S^{-}_{l}).
\end{eqnarray}

In the case $\delta_{l}\gg\lambda_{l}/2$, there is no energy
exchange between the artificial atomic system and the resonator.
The effective Hamiltonian is given by
\begin{eqnarray}\label{14}
H^{\mathrm{new}}_{\mathrm{eff}}&=&\sum_{l=1}^{4}\frac{(\lambda_{l})^{2}}{4\delta_{l}}(|s_{l}\rangle\langle
s_{l}|+|g_{l}\rangle\langle g_{l}|)\cr &&+\sum^{4}_{l,m=1(l\neq
m)}\frac{\lambda_{l}\lambda_{m}}{4}(\delta_{l}^{-1}+\delta_{m}^{-1})\cr&&(e^{-i(\delta_{l}-\delta_{m})t}S^{+}_{l}S^{+}_{m}+e^{-i(\delta_{l}-\delta_{m})t}S^{+}_{l}S^{-}_{m}\cr&&+\rm
H.c).
\end{eqnarray}

\section*{3. Generation of the  $\chi$-type four-particle entangled state}

In this section, we will show how to prepare the $\chi$-type
four-particle entangled state. In the strong driving regime,
\emph{i.e.} 2$\Omega_{S}\gg\delta_{l}, \lambda_{l},\eta_{l},\xi$,
the evolution operator of the system is
\begin{eqnarray}\label{16}
U(t)&=&e^{-iH^{\mathrm{new}}_{\mathrm{0}}t}e^{-iH^{\mathrm{new}}_{\mathrm{eff}}t},\nonumber\\
H^{\mathrm{new}}_{\mathrm{0}}&=&\sum^{4}_{l=1}2\Omega_{S}\sigma_{z,l},\nonumber\\
H^{\mathrm{new}}_{\mathrm{eff}}&=&\sum_{l=1}^{4}\alpha(|s_{l}\rangle\langle
s_{l}|+|g_{l}\rangle\langle g_{l}|)\cr &&+\sum^{4}_{l,m=1(l\neq
m)}\beta(e^{-i(\delta_{l}-\delta_{m})t}S^{+}_{l}S^{+}_{m}\cr&&+e^{-i(\delta_{l}-\delta_{m})t}S^{+}_{l}S^{-}_{m}+\rm
H.c),
\end{eqnarray}
where $\alpha=\frac{(\lambda_{l})^{2}}{4\delta_{l}},$
 $\beta=\frac{\lambda_{l}\lambda_{m}}{4}(\delta_{l}^{-1}+\delta_{m}^{-1})$.

Through a suitable choice of the Rabi frequencies and detunings of
the classical fields, we can use this system to carry out the
selective coupling between arbitrarily two qubits. Firstly, we
apply classical pulses to the AAs in TLR simultaneously. Among
them, we let $\Omega_{1}$$=$$\Omega_{2}$$=$$\Omega$,
$\Omega_{3}$$=$$\Omega_{4}$$=$$\Omega^{'}$,
 $\Delta_{1,1}$$=$$\Delta_{1,2}$$=$$\Delta_{1},$ $\Delta_{1,3}$$=$$\Delta_{1,4}$$=$$\Delta_{1}^{'}$
 and  $|\Delta_{1}-\Delta_{1}^{'}|\gg|\beta_{a,b}|$ $(a$$=$$1,2
 $ and $b$$=$$3,4)$. In this case, AA $1(3)$ is only coupled to $2(4)$,
 while $1$ and $2$ are decoupled to AAs $3$, $4$. The evolution operator of
the system is
\begin{eqnarray}\label{16}
U(t)&=&e^{-iH^{\mathrm{new}}_{\mathrm{0}}t}e^{-iH^{\mathrm{new}}_{\mathrm{eff}}t},\nonumber\\
H^{\mathrm{new}}_{\mathrm{0}}&=&\sum^{4}_{l}2\Omega_{S}\sigma_{z,l},\nonumber\\
H^{\mathrm{new}}_{\mathrm{eff}}&=&\alpha\sum^{2}_{l=1}(|s_{l}\rangle\langle
s_{l}|+|g_{l}\rangle\langle g_{l}|)\cr
&&+\alpha^{'}\sum^{4}_{l=3}(|s_{l}\rangle\langle
s_{l}|+|g_{l}\rangle\langle g_{l}|)\cr &&+\beta(
S^{+}_{1}S^{+}_{2}+S^{+}_{1}S^{-}_{2}+\rm H.c)\cr
&&+\beta^{'}(S^{+}_{3}S^{+}_{4}+S^{+}_{3}S^{-}_{4}+\rm H.c),
\end{eqnarray}
where $\alpha^{'}$ and $\beta^{'}$ have the same formulas as
$\alpha$ and $\beta$ with parameter values $\Omega^{'}$ and
$\Delta^{'}$ instead of $\Omega$ and $\Delta$ respectively.

Assume that the four AAs are initially in the state
$|gggg\rangle_{1234}$. After an interaction time $t_{1}$ the state
of the system is
\begin{eqnarray}\label{16}
&|gggg\rangle_{1234}&\rightarrow e^{-i2\alpha
t_{1}}e^{-i2\alpha^{'}t_{1}}\cr&&\{\cos(\beta
t_{1})[\cos(\Omega_{S}t_{1})|g_{1}\rangle-i\sin(\Omega_{S}t_{1})|s_{1}\rangle]\cr
&&[\cos(\Omega_{S}t_{1})|g_{2}\rangle
-i\sin(\Omega_{S}t_{1})|s_{2}\rangle]\cr &&-i\sin(\beta
t_{1})[\cos(\Omega_{S}t_{1})|s_{1}\rangle-i\sin(\Omega_{S}t_{1})|g_{1}\rangle]\cr
&&[\cos(\Omega_{S}t_{1})|s_{2}\rangle-i\sin(\Omega_{S}t_{1})|g_{2}\rangle]\}\cr
&&\{\cos(\beta^{'}t_{1})[\cos(\Omega_{S}t_{1})|g_{3}\rangle-i\sin(\Omega_{S}t_{1})|s_{3}\rangle]\cr
&&[\cos(\Omega_{S}t_{1})|g_{4}\rangle
-i\sin(\Omega_{S}t_{1})|s_{4}\rangle]\cr
&&-i\sin(\beta^{'}t_{1})[\cos(\Omega_{S}t_{1})|s_{3}\rangle-i\sin(\Omega_{S}t_{1})|g_{3}\rangle]\cr
&&[\cos(\Omega_{S}t_{1})|s_{4}\rangle-i\sin(\Omega_{S}t_{1})|g_{4}\rangle]\}.
\end{eqnarray}

The common phase factor $e^{-i2\alpha
t_{1}}e^{-i2\alpha^{'}t_{1}}$ can be discarded. When the
interaction time $t_{1}$ satisfies $\Omega_{S}t_{1}$$=$$n\pi$ ($n$
is an integer) and $\beta t_{1}$$=$$\beta^{'}t_{1}$$=$$\pi/4$, the
four-atom state becomes
\begin{eqnarray}\label{16}
|gggg\rangle_{1234}\rightarrow\frac{1}{2}(|gggg\rangle-i|ggss\rangle-i|ssgg\rangle-|ssss\rangle)_{1234}.
\end{eqnarray}

Then, we let $\Omega_{2}$$=$$\Omega_{3}$$=$$\Omega$,
$\Omega_{1}$$=$$\Omega_{4}$$=$$0$,
 and $\Delta_{1,2}$$=$$\Delta_{1,3}$$=$$\Delta_{1}$. And we only apply two
strong classical pulses to the AA $2$ and AA $3$ simultaneously.
In this step, only the interaction between AA
 $2$ and AA $3$ can happen. After the operation
 time $t_{2}$ under the condition $\Omega_{S}t_{2}$$=$$n\pi$,
 $\beta t_{2}$$=$$\pi/4$, we obtain an entangled state
\begin{eqnarray}\label{16}
|\chi^{00}\rangle^{'}_{1234}=\frac{1}{2\sqrt{2}}(|gggg\rangle-i|gssg\rangle-i|ggss\rangle-|gsgs\rangle\cr
-i|ssgg\rangle-|sgsg\rangle-|ssss\rangle+i|sggs\rangle)_{1234}.
\end{eqnarray}

Through a local unitary transformation on AAs $1$ and $3$
$\{|g\rangle\rightarrow|g\rangle$, $|s\rangle\rightarrow
i|s\rangle\}$, we can obtain the genuine four-qubit $\chi$-type
entanglement state
\begin{eqnarray}\label{16}
|\chi^{00}\rangle_{3214}=\frac{1}{2\sqrt{2}}(|gggg\rangle-|ggss\rangle-|gsgs\rangle+|sggs\rangle\cr
+|ssgg\rangle+|gssg\rangle+|ssss\rangle+|sgsg\rangle)_{3214}.
\end{eqnarray}

\section*{4. Discussion and conclusion}
\begin{figure}[htp]
\begin{center}
  \includegraphics[scale=0.7]{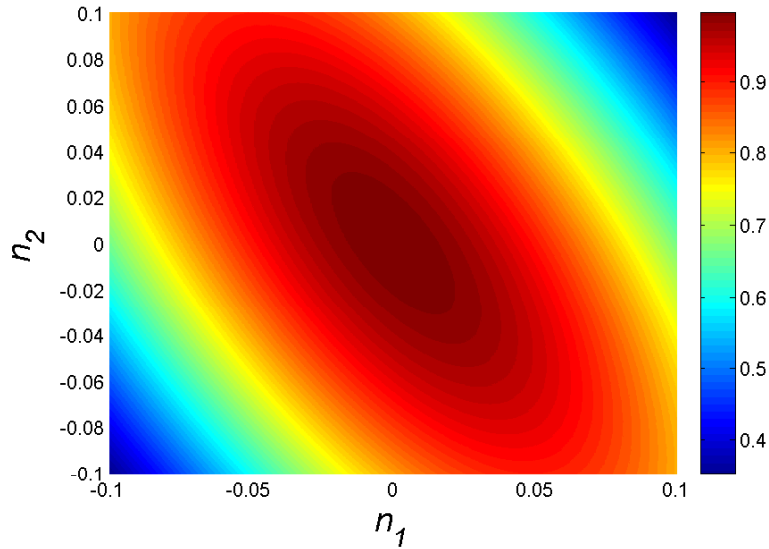}\\
  \caption{Density plot of the fidelity as a function of time error
 rates
$n_{1}$ and $n_{2}$.
  }\label{fig3}
  \end{center}
\end{figure}

In the above part, we have assumed the interacting times can be
controlled accurately. But in the realistic experiment, the
interacting time of each step can't be controlled perfectly, and
the time errors will inevitably exist. For the sake of
convenience, we define the time error rate as $n_{i}$$=$$\Delta
t_{i}/t_{i} (i= 1,2)$, $\Delta t_{i}$ is the time error in the
$i$th step. Considering time errors, we rethink the process of
derivation and get a real state $|\chi\rangle_{\mathrm{real}}$.
The fidelity is defined as
\begin{eqnarray}\label{16}
F=|_{3214}\langle \chi^{00}|\chi\rangle_{\mathrm{real}}|^{2}.
\end{eqnarray}
For an intuitive grasp of the effect of the time errors on the
fidelity, the total fidelity as function of $n_{i}$ is given in
Fig.~\ref{fig3}. It can be seen from Fig.~\ref{fig3} that the
fidelity decreases slightly with the increase of the interaction
time errors. Under the condition $n_{1}$$=$$0.02,n_{2}$$=$$0.02$,
the fidelity can still reach 0.96.

 Now we briefly discuss the feasibility of the present
scheme based on the current available parameters. At the first
step, we set $\Omega_{1}$$=$$\Omega_{2}$$=$$\Omega$$=$$g$,
$\Omega_{3}$$=$$\Omega_{4}$$=$$\Omega^{'}$$=$$0.725g$,
$\Delta_{1,1}$$=$$\Delta_{1,2}$$=$$10g$,
$\Delta_{1,3}$$=$$\Delta_{1,4}$$=$$10.5g$, and
$\Delta_{2}$$=$$11g$. Then we have
$\beta\simeq\beta^{'}\simeq4.56\times10^{-3}g$, so the time of the
first step is $t_{1}$$=$$\pi/4\beta\simeq0.172\times10^{3}/g$. At
the second step, we set
$\Omega_{2}$$=$$\Omega_{3}$$=$$\Omega$$=$$g$,
$\Omega_{1}$$=$$\Omega_{4}$$=$$0$,
$\Delta_{1,2}$$=$$\Delta_{1,3}$$=$$\Delta$$=$$10g$ and
$\Delta_{2}$$=$$11g$. Then we have
$\beta\simeq4.56\times10^{-3}g$, so the time of the second step is
$t_{2}$$=$$\pi/4\beta\simeq0.172\times10^{3}/g$. It is known that,
with current experimental techniques, the coupling constant
between the flux qubit and the data bus can reach
$g$$=$$2\pi\times200 MHz$\cite{LYX2006}. Therefore, in the first
step, the operation time is about $t_{1}$$=$$0.137\mu s$. And in
the second step, the parameters remain unchanged, then the
operation time $t_{2}$ is the same as $t_{1}$. The total operation
time is $T$$=$$t_{1}+t_{2}$$=$$0.274\mu s$. However, in the
experiment the relaxation time $\tau_{r}$ and dephasing time
$\tau_{d}$ of the flux qubit can reach
$\tau_{r}\sim\tau_{d}$$=$$1.5\mu s$\cite{LP2010}. Thus, it is
obvious that $T=t_{1}+t_{2}\ll\tau_{r}$.

In conclusion, we have proposed a new scheme to create a genuine
$\chi$-type four-particle entangled state of superconducting AAs
in TLR with broken symmetry, in which selection rules can not hold
and cyclic transition structures are generated. The distinguished
feature of the present scheme is that artificial atomic excited
states and photonic states are adiabatically eliminated, so our
scheme is robust against the spontaneous emissions of AAs and the
decays of TLR. We have analyzed the performance and the
experimental feasibility of the scheme, and shows that our scheme
is feasible under existing experimental conditions.

\begin{center}$\mathbf{Acknowledgments}$\end{center}
This work was supported by the National Natural Science Foundation
of China under Grant Nos. 11464046 and 61465013.

\end{document}